\documentclass[12pt]{JHEP3} 
\pdfoutput=1

\usepackage{graphicx}
\usepackage[numbers,sort&compress]{natbib}
\usepackage{amsmath}
\usepackage{amssymb}
 
\def\beq{\begin{equation}}
\def\eeq{\end{equation}}
\def\beqa{\begin{eqnarray}}
\def\eeqa{\end{eqnarray}}
\def\ss{\scriptscriptstyle}
\newcommand\tr{\mathop{\mathrm{Tr}}}

\makeatletter
\renewcommand\@ENVwarn[1]{}
\makeatother

\title{A non-homogeneous ground state of the low-temperature Sakai-Sugimoto model}

\author{C.A.~Ballon Bayona, Kasper Peeters and Marija Zamaklar \\
Department of Mathematical Sciences,\\
Durham University,\\
South Road,\\
Durham DH1 3LE,\\
United Kingdom.\\
\email{c.a.m.ballonbayona@durham.ac.uk} \\
\email{kasper.peeters@durham.ac.uk} \\
\email{marija.zamaklar@durham.ac.uk}}

\date{April 11th, 2011}
\preprint{DCPT-11/15}

\abstract{We study the instability of the low-temperature QCD vacuum
  at \mbox{large-$N_c$} due to an axial chemical potential, using the
  holographic Sakai-Sugimoto model.  We explicitly construct the
  ground state of the theory, which corresponds to a translationally
  non-invariant configuration.}

\begin{document}

\section{Introduction}

The behaviour of QCD at non-zero temperature and density is an active
area of research both from a theoretical as well as an experimental
perspective. While perturbation theory can be used at very large
temperatures or densities, this is of limited use in intermediate
regimes such as those that are relevant for heavy ion collisions or
the description of the interior of neutron stars, where QCD is
strongly coupled. Lattice methods in turn are hard to apply to highly
time-dependent processes, and also pose difficulties when dealing with
systems with large baryon density because of the sign
problem. Alternative techniques are therefore called for. The
gauge/gravity duality provides a non-conventional approach to the
study of strongly coupled gauge theories, and while the dual of QCD
proper is not known (not even in the large-$N_c$ limit), some of the
current proposals for holographic duals of gauge theories do at least
seem to capture several qualitative properties of strongly coupled
QCD.

One of the key questions related to QCD at non-zero quark chemical
potential is that of the character of the ground state. Various models
suggest that, for particular values of the chemical potential and
temperature, an instability sets in which results in the decay of the
homogeneous vacuum to one which breaks translational invariance (see
for instance the ladder analysis of QCD in the large-$N_c$ limit by
Deryagin, Grigoriev and Rubakov~\cite{Deryagin:1992rw}). In the dual
holographic language, chemical potentials are typically realised by
introducing an electric gauge field configuration on the world-volume
of a flavour brane, with the non-normalisable component being related
to the value of the chemical potential. Such electric configurations
in the presence of a curved background can become unstable, as was
emphasised in work of Nakamura, Park and Ooguri
\cite{Nakamura:2009tf}.  These authors have pointed out that a black
hole with a constant electric field in five-dimensional
Maxwell-Chern-Simons theory is unstable against decay towards a
configuration with both electric and magnetic fields turned
on. Importantly, this instability occurs only for a non-vanishing
value of the spatial momentum~$k$, signalling that the new ground
state is spatially modulated.  The lowest energy state has a uniquely
specified momentum $k_{\text{gs}}$, which is determined by the value
of the electric field.  The origin of the instability can be traced to
the presence of a Chern-Simons coupling.

In the context of holographic models, such behaviour was in fact first
found in the work of Harvey and Domokos~\cite{Domokos:2007kt}, who
exhibited an instability in a particular bottom-up model with a
Chern-Simons coupling and an axial chemical potential. More recently,
Ooguri and Park~\cite{Ooguri:2010kt,Ooguri:2010xs} have analysed the
Sakai-Sugimoto model~\cite{Sakai:2004cn,Sakai:2005yt} in the
\emph{deconfined} phase, where chiral symmetry is restored. This model
contains Chern-Simons couplings in the action of flavour
D8-branes. The presence of a non-vanishing electric field
corresponding to a quark chemical potential then leads to an
instability towards a phase where baryonic charge density is
generated. This non-homogeneous ground state contains corresponding
baryonic and axial currents carrying non-zero momenta~$k_{\text{gs}}$.

\medskip

This still leaves open the question as to whether a similar
instability can occur at vanishing temperature. In the aforementioned
Sakai-Sugimoto model, it is known that even at zero temperature, a
sufficiently large \emph{isospin} chemical potential can lead to an
instability towards a new ground state in which vector mesons
condense~\cite{Aharony:2007uu}.  This ground state, even though it
breaks rotational invariance, is spatially homogeneous. The question
remains whether a non-homogeneous ground state is possible in this
model.  In the present paper, we therefore set out to use the
holographic model of Sakai and Sugimoto to study the properties of a
QCD-like theory in the large-$N_c$ limit, at non-vanishing values of
the \emph{axial} chemical potential.\footnote{An axial chemical
  potential is induced by a time-dependent theta angle, and as such is
  one of the motivations to study the Chiral Magnetic
  Effect~\cite{Fukushima:2008xe}, where theta angle fluctuations
  together with a magnetic field induce an electric current. The axial
  potential also plays an important role in the holographic
  realisation of chiral magnetic spirals~\cite{Kim:2010pu}. In the
  present paper, however, we do not turn on any additional magnetic
  field but simply study the consequence of the axial chemical
  potential by itself.}  More precisely, we analyse the ground state
in the \emph{confining} phase where \emph{chiral symmetry is
  broken}. Although the axial currents are anomalous in real QCD, this
anomaly is suppressed in the large-$N_c$ limit, and the axial $U(1)$
current is conserved. We will see that this potential can trigger an
instability to a new non-homogeneous ground state. This is to be
contrasted with the effect of a baryon chemical potential. The latter
would lead to condensation of baryons, which are somewhat problematic
in the large-$N_c$ limit as they become infinitely heavy, and are in
fact singular objects in the Sakai-Sugimoto model.\footnote{The
  authors of \cite{Chuang:2010ku} found a dynamical instability
  treating the baryonic chemical potential as a source of cusps in the
  flavour D8-branes.} Our solutions are regular and have finite
energy.\footnote{Because the solutions originate from the Chern-Simons
  term, they are of course still such that the fields scale as $\sim
  \lambda$, i.e.~their scale is set by the string length.}

Our results show that even the low-temperature phase of the
Sakai-Sugimoto is unstable against decay into a non-homogeneous
state. These results can be contrasted with recent
work~\cite{Chernodub:2011fr} on the effects of the axial chemical
potential in the ${\rm PLSM}_q$ model, where no evidence for an axial
potential induced phase transition at zero temperature was found. The
spatial modulation of the ground state we find is characterised by the
momentum, which is interestingly, to good approximation independent
of the axial density (or chemical potential). We also find that the
axial density of the new ground state is substantially larger for a given
chemical potential than in the homogeneous phase. The instabilities
towards new ground state occur both in the Maxwell-Chern-Simons
truncation of the flavour D8-brane action as well as in the full DBI
system, provided the Chern-Simons coupling of the latter is
sufficiently large. We do not find any evidence for condensation into
other states as the chemical potential is increased further. While new
non-homogeneous configurations appear, they do not correspond to new
global energy minima, but only to new meta-stable vacua, as they have larger
free energy than the true ground state.

\section{Chemical potentials in the Sakai-Sugimoto model}
\subsection{Review of the model}

In this section we briefly review the basics of Sakai-Sugimoto model
\cite{Sakai:2004cn,Sakai:2005yt} in order to set the scene and to
introduce the conventions we will be using in this paper. The
Sakai-Sugimoto model at low temperature is derived by considering the
decoupling limit of a large number $N_c$ of D4-branes compactified on
a circle of radius~$R$ spanned by the coordinate~$\tau$. Anti-periodic
boundary conditions are imposed on the fermions. The system
furthermore includes $N_f$ flavour D8-branes positioned at $\tau=0$
and $N_f$ anti-D8-branes positioned at $\tau=L$. The dual gauge theory
is a maximally supersymmetric $SU(N_c)$ gauge theory in 4+1
dimensions, which is compactified on a circle with anti-periodic
boundary conditions for the adjoint fermions, coupled to $N_f$
left-handed fermions in the fundamental representation of $SU(N_c)$
localised at~\mbox{$\tau=0$} and $N_f$~right-handed fermions in the
fundamental representation localised at~\mbox{$\tau=L$}.
 
In more detail, the background generated by the stack of D4-branes has
a metric, Ramond-Ramond four-form and string coupling given by 
\begin{equation}
\label{e:D4background}
\begin{aligned}
{\rm d}s^2 =&
\frac{U^{3/2}}{R^{3/2}} \, \eta_{\mu \nu} {\rm d}x^\mu {\rm d}x^\nu +
\frac{U^{3/2}}{R^{3/2}} f(U) {\rm d}\tau^2 + \frac{R^{3/2}}{U^{3/2}}
\frac{{\rm d} U^2}{f(U)} + R^{3/2} U^{1/2} {\rm d} \Omega_4^2 \, ,\\[1ex]
e^\phi =& g_s \frac{U^{3/4}}{R^{3/4}}\,, \quad F_4 = \frac{(2 \pi l_s)^3
  N_c}{V_{\ss S^4}} \epsilon_4\,, \quad f(U) = 1 - \frac{U_{KK}^3}{U^3} \, ,
\end{aligned}
\end{equation}
where $\epsilon_4$ is the volume form on $S^4$ and $V_{\ss S^4}$ its
volume. The index $\mu=0,1,2,3$ defines the directions of the dual
gauge theory, and $\tau$ is a coordinate on the circle so that
$\tau\equiv \tau + \delta\tau$. The curvature radius is given by $R =
(\pi g_s N_c)^{1/3} l_s$ where $g_s$ is the string coupling and $l_s
=\sqrt{\alpha'}$ is the string length. The period of the compact
coordinate introduces a four-dimensional mass scale
\begin{equation}
\label{periodic}
\delta \tau = \frac{4\pi}{3} \frac{R^{3/2}}{U_{\ss KK}^{1/2}} \equiv \frac{2 \pi}{M_{\ss KK}} \, , 
\end{equation}
and the four dimensional 't Hooft constant is related to the string
coupling by 
\begin{equation}
\lambda = g_{YM}^2 N_c = 2 \pi
M_{\ss KK} g_s N_c \, l_s \,.  
\end{equation}
For our purposes it will also be convenient to introduce use pair of
dimensionless coordinates $y$ and $z$ defined by the relations
\begin{equation}
U = U_{\ss KK} \left (1 + y^2 + z^2 \right )^{1/3} \equiv U_{\ss KK}
K_{y,z}^{1/3} \quad , \quad \tau = \frac{\delta \tau}{2 \pi} \arctan
\left (\frac{z}{y} \right) \,.
\end{equation} 
In terms of these coordinates the metric takes the form
\begin{multline}
{\rm d}s^2 = U_{\ss KK}^{3/2} R^{-3/2}
K_{y,z}^{1/2} \, \eta_{\mu \nu} {\rm d}x^\mu {\rm d}x^\nu \,+\, \frac49 R^{3/2}
U_{\ss KK}^{1/2} \frac{K_{y,z}^{-5/6}}{y^2 + z^2} \Big [ (z^2 + y^2
K_{y,z}^{1/3}){\rm d}z^2\\[1ex]
+ (y^2 + z^2 K_{y,z}^{1/3} ) {\rm d}y^2 + 2 y z (1
- K_{y,z}^{1/3}) {\rm d}y {\rm d}z \Big ] + R^{3/2} U_{\ss KK}^{1/2} K_{y,z}^{1/6}
{\rm d}\Omega_4^2 \,.  
\end{multline}

In this background one introduces a probe D8-brane and $\overline{{\rm
    D}8}$-brane which fill out the $3+1$ dimensional space-time and
the four-sphere, and trace out a curve in the remaining two directions
$(U,\tau)$. The latter coordinates span a two dimensional cigar-like
subspace.  The shape of the curve is obtained by solving the equation
of motion of the probe branes, and is given by a one parameter family,
which is specified by the asymptotic separation between the D8 and
$\overline{{\rm D}8}$-brane. In this paper we will be interested in
the so-called antipodal brane embedding corresponding to the maximal,
antipodal separation of the branes in the circular direction $\tau$. In
this case, the branes extend all the way to the tip of the cigar where
they join in a smooth way. For this special case the metric on the
brane world-volume takes the form
\begin{equation}
\label{gD8brane}
\begin{aligned}
 {\rm d}s_{D8}
&= U_{\ss KK}^{3/2} R^{-3/2} K_z^{1/2} \, \eta_{\mu \nu} {\rm d}x^\mu
{\rm d}x^\nu \,+\, \frac49 R^{3/2} U_{\ss KK}^{1/2} K_z^{-5/6} {\rm d}z^2 +
R^{3/2} U_{\ss KK}^{1/2} K_z^{1/6} {\rm d}\Omega_4^2 \\[1ex]
 &\equiv G_{M N}
{\rm d}x^M {\rm d}x^N \, , 
\end{aligned}
\end{equation}
where $K_z = 1 + z^2$. 

The action for the $U(N_f)$ gauge fields living on the
D8-$\overline{{\rm D}8}$-branes is given by the DBI action which in
the non-abelian limit is not fully known, but at the quadratic order
reduces just to the the Yang-Mills action in a curved background,
corresponding to the curved world-volume of the brane,
\begin{equation}
S_{\text{YM}} = - \mu_8
(\pi \alpha ')^2 \int {\rm d}^4 x {\rm d}z {\rm d}^4 \Omega \, e^{-\phi } \sqrt{- |G| }
\tr{ \left ( {\cal F}^{MN} {\cal F}_{MN} \right )}
\end{equation} 
where ${\cal  F}_{MN} = \partial_M {\cal A}_N -
\partial_N {\cal A}_M + i [{\cal A}_M , {\cal A}_N]$, and the index
$m=(\mu,z, \alpha)$. The coordinate $z$ spans the holographic direction
and $\alpha$ labels the directions on $S^4$; indices are raised or
lowered using the metric (\ref{gD8brane}).  In order to avoid
non-singlet $SO(5)$ states associated to the $S^4$, in what follows we
will set all the excitations along the sphere directions to zero,
${\cal A}_\alpha=0$ and assume that ${\cal A}_\mu$ and ${\cal A}_z$ do
not depend on the $S^4$ coordinates.  Then we can integrate out the
$S^4$ coordinates and the Yang-Mills action reduces to
\begin{equation}
 S_{\text{YM}} =
-\kappa\int {\rm d}^4 x {\rm d} z \tr \left [ \frac12 K_z^{-1/3} \eta^{\mu \rho}
  \eta^{\nu \sigma} {\cal F}_{\mu \nu} {\cal F}_{\rho \sigma} + M_{\ss
    KK}^2 K_z \eta^{\mu \nu} {\cal F}_{\mu z} {\cal F}_{\nu z} \right
]
\end{equation} 
where 
\begin{equation} 
\kappa= \frac{\sqrt{\alpha '} g_s N_c^2 M_{\ss
    KK}}{108 \pi^2} = \frac{\lambda N_c}{216 \pi^3} \,.  
\end{equation} 
Note that the Yang-Mills action can now be written in terms of an
effective five-dimensional metric 
\begin{equation} 
{\rm d}s_{(5)}^2 = g_{mn} {\rm d}x^m {\rm d}x^n =
M_{\ss KK}^2 K_z^{2/3} \, \eta_{\mu \nu} {\rm d}x^\mu {\rm d}x^\nu \,+\,
K_z^{-2/3} {\rm d}z^2 \, , \label{gmn}
\end{equation}
in terms of which it takes the form
\begin{equation} 
S_{\text{YM}} = - \frac{\kappa}{2} \int\!{\rm d}^4x {\rm d}z 
 \sqrt{-g} \tr \left [ {\cal F}^{mn} {\cal F}_{mn} \right] \, ,
\end{equation}
where the indices are now raised or lowered using the metric $g_{mn}$.

The effective action for the D8 brane includes also a Chern-Simons
term which is given by 
\begin{equation} 
\label{e:SCS}
S_{\text{CS}} = \mu_8 \frac{(2 \pi \alpha ')^3}{3 !} \int_{D8}\! \omega_5 \wedge F_4 = \alpha\int\! \omega_5
\, , \qquad \alpha= \hat{\alpha}\frac{N_c}{24 \pi^2}\,.
\end{equation} 
Here $\hat{\alpha}=1$ in string theory but we will keep it as a
separate parameter for later use\footnote{For comparison, our
  $\hat\alpha$ corresponds to $4/3$ times $\alpha$ used
  in~\cite{Ooguri:2010xs}.}.  Again, $F_4$ is the Ramond-Ramond
four-form, proportional to the volume form of $S^4$
(see~\eqref{e:D4background}), and the $\omega_5$ form is the usual
five-dimensional Chern-Simons form,
\begin{equation} 
\omega_5 = \tr{ \left ( {\cal A} {\cal F}^2 -
    \frac{i}{2} {\cal A}^3 {\cal F} - \frac{1}{10} {\cal A}^5 \right )
} \,.  
\end{equation} 
It is useful for our purposes to decompose the $U(N_f)$ gauge field
into the $SU(N_f)$ part $\tilde{{\cal A}}$ and the $U(1)$ part
$\hat{{\cal A}}$ as
\begin{equation}
{\cal A} = \tilde{{\cal A}} + \frac{1}{\sqrt{2 N_f}} \hat{\cal{A}}
\end{equation}
For the case  $N_f=2$, we can decompose the $U(2)$ symmetry into the
baryon $U(1)$ and isospin $SU(2)$, and the Chern-Simons term
decomposes as
\begin{equation}
S_{\text{CS}} = \alpha \int \bigg{(} \frac{3}{2} \hat{{\cal A}} \tr
{\cal F}^2
 + \frac{1}{4} \hat{{\cal A}}\hat{{\cal F}^2} + (\text{total derivatives}) \bigg{)}\,,
\end{equation}
We see that in the case of two D-branes, the Chern-Simons coupling has
two terms, one involving only $U(1)$ fields and the other one which
couples the $U(1)$ and the $SU(2)$ fields. In the present paper we
will focus on the effect of the $U(1)$ part.

\subsection{Holographic realisation of chemical potentials}

Holographic models encode \emph{global} symmetries of the dual gauge
theory in the form of \emph{gauge} symmetries in the bulk theory, and
these relations hold both for the closed as well as the open sectors
of the string theory. So in particular, in the D8-$\overline{{\rm
    D}8}$ system, there are two independent gauge theories living near
the two boundaries of the flavour brane system, giving rise to 
$U(N_f)_L$ and $U(N_f)_R$ gauge symmetries, and these correspond to
the global $U(N_f)\times U(N_f)_R$ of the dual gauge theory.

In the low-temperature phase of the Sakai-Sugimoto model that we are
interested in, the two branes are connected in the interior of the
bulk space, and thus the gauge fields ${\cal A}_{M}^L$ and ${\cal A}_{M}^R$ are
limits of a single gauge field living on the two connected
branes. Therefore, one cannot independently perform gauge
transformations on these two gauge fields, but is constrained to gauge
transformations which, as $z\rightarrow \pm \infty$, act in a related
way. Specifically, since near the boundary a large bulk gauge
transformation acts as ${\cal A}_{L/R} \rightarrow g_{L/R} {\cal A}_{L/R} g^{-1}_{L/R}
$, then clearly if $g_{L} = g_{R}$ any state (and in particular the
trivial vacuum ${\cal A}=0$) is invariant under the vectorial transformations,
i.e. those transformations for which $g_{L} = g_{R}$.  Hence it means
that the vector-like symmetry is unbroken in this model.

The correspondence between the bulk gauge field and the source and
global symmetry current of the dual gauge theory is encoded in the
asymptotic behaviour of the former. More precisely, the bulk gauge
field ${\cal A}_M(z,x^\mu)$ behaves, near the boundary and in the
${\cal A}_z=0$
gauge, as
\begin{equation}
\label{asymptotic}
{\cal A}_n(x^\mu,z) \rightarrow 
  a_{n}(x^\mu)\Big(1 + \mathcal{O}(z^{-2/3})\Big) 
   + \rho_n(x^\mu) \frac{1}{z}\Big(1 + \mathcal{O}(z^{-2/3}) \Big)\,.
\end{equation}
Here $a_n(x^\mu)$ describes non-normalisable behaviour of the field,
and is interpreted as a source in a dual field theory action of
the form
\begin{equation}
\int\!{\rm d}^4x\, a_n(x^\mu) J^n(x^\mu)\,,
\end{equation}
while $\rho_n(x^{\mu})$ is proportional to the expectation value of
the current $J_n(x^\mu)$. Therefore, adding a chemical potential to
the field theory corresponds to adding a source for $J^0$, which
implies the boundary condition for the holographic gauge field
${\cal A}_n(x)= \mu \delta_{n0}$.

For the Sakai-Sugimoto model, the bulk field ${\cal A}_m$ living on the
D8-branes has \emph{two} asymptotic regions, corresponding to each
brane, and hence there are two independent chemical potentials $\mu_L$
and $\mu_R$ which can be separately turned on. Instead of left and
right chemical potentials one often introduces vectorial and axial
potentials, defined respectively as $\mu_V=\frac{1}{2}(\mu_L +\mu_R)$
and $\mu_A=\frac{1}{2}(\mu_L-\mu_R)$. The vectorial and axial chemical
potentials for the $U(1)$ subgroup of the $U(2)$ gauge group on the two
D8-branes correspond to the baryonic and axial chemical potential in the
dual gauge theory, while the non-abelian $SU(2)$ chemical potentials
are mapped to the vectorial and axial isospin potentials.

In this paper we will be interested only in \emph{axial} chemical
potentials. Our computations will mainly be concerned with the $U(1)$
subgroup, since the instability which
leads towards the spatially modulated phase in the $U(1)$ subgroup is
present in the same form in the $U(1)$ subgroup of an isospin $SU(2)$
group, leading to the instability of homogeneous configuration for the
isospin chemical potential as well.

Our starting point is the \emph{homogeneous}
solution~\cite{Sakai:2004cn} in the ${\cal A}_z=0$ gauge. For the configuration
to describe an axial chemical potential one needs to turn on only
the ${\cal A}_0(z)$ component, and look for odd, non-normalisable solutions to
the equations of motion (\ref{abelianEOM}). It is not hard to see that
with this simple ansatz the Chern-Simons contribution to the equations of
motion vanishes, and the solution to the equation of motion is given by
\begin{equation}
\label{solutionsimpl}
{\cal A}_0(z) = \frac{2}{\pi} \mu_A \arctan z\,.
\end{equation}
Just as in the chiral Lagrangian, there is no potential generated on
moduli space~\cite{Aharony:2007uu}, and the result of the axial
potential is simply a non-vanishing axial density. The value of this
density is easily read off using \eqref{asymptotic} to be $\rho_A =
2\mu_A/\pi$.  This same solution can also be trivially embedded in the
$U(1)$ subgroup of the $SU(2)$ isospin group. In this case however, as
explained in \cite{Aharony:2007uu}, the odd ${\cal A}_0(z)$ configuration in
the ${\cal A}_z=0$ gauge is equivalent (by a global $SU(2)$ rotation) to a
system with a vectorial chemical potential and a non-trivial pion
condensate.

\section{The spatially modulated phase}

We have seen in the previous section that turning on the axial
chemical potential in Sakai-Sugimoto model requires finding a
non-normalisable spatially homogeneous solution for the $U(1)$ field
${\cal A}_0$, which is an \emph{odd} function in the holographic
direction. The asymptotic values of ${\cal A}_0$ correspond to the values of
the chemical potentials in the dual gauge theory.

As mentioned in the introduction, the work of Domokos and
Harvey~\cite{Domokos:2007kt} and Nakamura, Ooguri and
Park~\cite{Nakamura:2009tf,Ooguri:2010kt} has shown that in Maxwell
theory with a Chern-Simons coupling turned on, a constant electric
field is \emph{unstable} against decay to a spatially modulated phase,
in which a magnetic field transverse to the direction of the initial
electric field is switched on. Since the configuration with
non-vanishing axial chemical potential~\eqref{solutionsimpl} amounts
to having a non-vanishing electric field in the bulk, we expect that
an instability similar to that found by the aforementioned authors
should be present here.  In the present section we will analyse the
Sakai-Sugimoto model at zero temperature with non-zero axial chemical
potential, show that for sufficiently large values of that potential a
new ground state appears, and determine that it has lower energy than
the homogeneous vacuum.

\subsection{Equations of motion}

Let us begin by writing the effective 5-dimensional action for the
case of a single flavour $N_f=1$, obtained after we integrate out the
fields on the four sphere $S^4$, for the case of the single
D8-$\overline{{\rm D}8}$ brane,
\begin{equation}
 S_{\text{YM}} + S_{\text{CS}} =
- \frac{\kappa}{2} \int\! {\rm d}^4 x {\rm d}z \sqrt{-g} \, {\cal F}^{mn} {\cal
  F}_{mn} + \frac{\alpha}{4} \, \epsilon^{\ell m n p q} \int\! {\rm
  d}^4x {\rm d}z {\cal A}_{\ell} {\cal F}_{mn} {\cal F}_{pq} \, .\label{YMCS} 
\end{equation}
The equations of motion following from this action read
\begin{equation}
\begin{aligned}
\label{abelianEOM}
&\sqrt{- g} \left [ g^{00} g^{zz} \partial_0 {\cal F}_{0 z}
  \,+\, g^{xx} g^{zz} \partial_i {\cal F}_{i z} \right ] + \frac32
\frac{\alpha}{\kappa} \epsilon^{i j k} {\cal F}_{0 i} {\cal F}_{j k}
\,=\, 0 \, , \\[1ex]
&\partial_z \left [ \sqrt{- g} \, g^{zz} g^{00}
  {\cal F}_{z 0} \right ] + \sqrt{- g} \, g^{00} g^{xx} \partial_i
      {\cal F}_{i 0} - \frac32 \frac{\alpha}{ \kappa} \epsilon^{i j k}
      {\cal F}_{z i} {\cal F}_{j k} \,=\, 0 \, , \\[1ex]
&\partial_z
      \left [ \sqrt{- g} \, g^{zz} g^{xx} {\cal F}_{z i} \right ] +
      \sqrt{- g} \, (g^{xx})^2 \partial_j {\cal F}_{j i} + \frac32
      \frac{\alpha}{ \kappa} \epsilon^{i j k} {\cal F}_{z 0} {\cal
        F}_{j k} -\frac{3\,\alpha}{ \kappa} \epsilon^{i j k} {\cal
        F}_{z j} {\cal F}_{0 k} \,=\, 0 \, .
\end{aligned}
\end{equation}
Here we have split the five-dimensional bulk indices as
$m=(i=(1,2,3),0,z)$ and we have assumed that the general diagonal
five-dimensional effective metric $g_{mn}$ depends only on the $z$
coordinate and has symmetry $g_{11}=g_{22}=g_{33}=g_{xx}$ (valid for
the Sakai-Sugimoto system).

It is important to note that the variation of the action \eqref{YMCS}
leads to a boundary term, which, after one uses the Bianchi identity
$\partial_m {\cal F}_{pq} + \partial_q {\cal F}_{m p} + \partial_p
{\cal F}_{q m} = 0$, reads
\begin{equation}
\label{boundaryeom}
\delta S_{\text{bdy}} = - \int\! {\rm d}^4x {\rm d}z \,
\partial_m \left [ \delta {\cal A}_\ell \left ( 2 \kappa\sqrt{-g} \, {\cal F}^{m \ell} 
+ \alpha\epsilon^{\ell m n p q}  {\cal A}_n {\cal F}_{pq}  \right )  \right ] \, .
\end{equation}
Therefore, when $\delta {\cal A}_l$ does not vanish, one has to add a
boundary term to the action in order to ensure that there are no
boundary contributions to the equations of motion~(see
e.g.~\cite{Braden:1990hw} for a discussion in related context). This
is important when we consider the system at constant density. In the
appendix we discuss this issue in more detail, and give the
corresponding boundary term for the antipodal embedding of
D8-$\overline{{\rm D}8}$ branes.

\subsection{The abelian ansatz}

To describe a spatially modulated phase, our starting point is the
ansatz
\begin{equation}
\label{ansatznonhom}
 {\cal A}_0 = f(z,\vec{x}) \,,\quad \vec{{\cal A}} =
\vec{\cal A} (z,\vec{x}) \, ,  
\end{equation}
where $\vec{{\cal A}}$ are in the spatial directions of the boundary.  We
will work in the gauge ${\cal A}_z=0$. Furthermore, we only want to
turn on the chemical potential for the axial currents and insist that
the spatially modulated phase is spontaneously generated (i.e.~without
introducing any kind of sources). Hence we impose normalisable
boundary conditions on the fields $\vec{{\cal A}}$, while similar to
the homogeneous case we impose that ${\cal A}_0$ is non-normalisable,
and tends to opposite constants at the two ends of the D8-brane
system,
\begin{equation}
\vec{{\cal A}}( z = \pm \infty ) \,=\, 0 \,, 
\quad \quad {\cal A}_0( z = \pm \infty ) = \pm \frac{2}\pi \mu_A \,.
\end{equation}
For this ansatz (\ref{ansatznonhom}) the full non-linear equations of
motion (\ref{abelianEOM}) become\footnote{In order to compare our
  normalisation of the Chern-Simons coupling $\alpha$ with
  \cite{Ooguri:2010xs}, note that equation (6) of that paper contains a typo: the
  coefficient $\alpha$ should read $\alpha/6$.}
\begin{align}
&\sqrt{-g} \, g^{xx} g^{zz} \partial_z ( \nabla \cdot \vec{\cal A}) - 3 \frac{\alpha}{ \kappa} \, \vec{\cal E} \cdot \vec{\cal B} \,=\,0 \, \label{geneq1}, \\[1ex]
&\partial_z \left [ \sqrt{- g} \, g^{zz} g^{00} \partial_z f \right ] - \sqrt{- g} \, g^{00} g^{xx} (\nabla \cdot \vec{\cal E}) - 3 \frac{\alpha}{ \kappa}  (\partial_z \vec{\cal A}) \cdot \vec{\cal B} \,=\, 0 \, , \label{geneq2} \\[1ex]
& \partial_z \left[ \sqrt{- g} \, g^{zz} g^{xx} \partial_z \vec{\cal A}  \right ] - \sqrt{- g} (g^{xx})^2 \nabla \times \vec{\cal B} +  3 \frac{\alpha}{ \kappa} \left [ (\partial_z f) \vec{\cal B} - \partial_z \vec{\cal A} \times \vec{\cal E} \right ] = 0 \,. \label{geneq3} 
\end{align}
In these equations, $\vec{\cal E} = {\cal F}_{0i} \, \hat x^i = -
\nabla f$ and $\vec{\cal B} = \frac12 \epsilon_{ijk} {\cal F}_{jk}
\hat x^i = \nabla \times \vec{\cal A}$ are the (bulk) transverse
electric and magnetic field associated to $f$ and $\vec{\cal A}$.

A simple ansatz that solves these equations is to impose 
\begin{equation}
\vec{\cal E} = 0 \,, \quad \vec{\cal B} \,=\,k \, \vec{\cal A} \, , \label{OPcond} 
\end{equation} 
where $k=\pm |\vec{k}|$ and $\vec{k}$ is the spatial momentum. This way the equation~\eqref{geneq1} is automatically satisfied and the equation~\eqref{geneq2} 
can be integrated.
The gauge field consistent with this ansatz can be written
as
\begin{equation}
\label{fandAsoln}
{\cal A}_0 = f(z)\,, \quad \vec{\cal A} \,=\, \vec{\eta}_1(z) \cos ( \vec{k}
\cdot \vec{x}) + \vec{\eta}_2 (z) \sin ( \vec{k} \cdot \vec{x}) \, ,
\end{equation}
where $\vec{\eta}_1$, $\vec{\eta}_2$ are polarisation vectors
satisfying 
\begin{equation}
\vec{\eta}_1 \times \vec{k} = k \, \vec{\eta}_2 \,, 
\quad \vec{k} \times \vec{\eta}_2 = k \, \vec{\eta}_1 \,.  
\end{equation}
For simplicity we choose  $\vec{k}=k \, \hat  x_1 $ so that the polarisation vectors reduce to $\vec{\eta}_1 (z) = h(z) \hat x_2$,  $\vec{\eta}_2 (z) =   - h(z) \hat x_3$ 
where $h(z)$ is a function that vanishes at  $z \to \pm \infty$.  The equations~\eqref{geneq2} and~\eqref{geneq3} take the form
\begin{equation}
\label{equationf}
M_{\ss KK}^2 K_z \partial_z f =
 \tilde \rho -
  \frac32 \frac{\alpha}{ \kappa} \, k \, h^2  \,,
\end{equation}
\begin{equation}
\label{hequation}
M_{\ss KK}^4 K_z \partial_z \left[ K_z \partial_z h \right ] - M_{\ss KK}^2 K_z^{2/3} k^2 h + 3 \frac{\alpha}{ \kappa} k h  \left[\tilde \rho  -  \frac32 \frac{\alpha}{ \kappa} \, k h^2 \right ] \,=\,0 \, ,
\end{equation}
where $\tilde \rho$ is an integration constant, $K_z=1 + z^2$ and $M_{kk}$ is defined in (\ref{periodic}). Equations~\eqref{equationf} and~\eqref{hequation} 
form the system that determines the new vacuum. In order to solve numerically this system it is convenient to introduce dimensionless variables 
$\hat f$, $\hat h$, $\hat k$ and $\hat \rho$ defined by
\begin{equation}
f =  \bar \lambda \, M_{\ss KK} \hat f \quad \, \quad 
h = \bar \lambda \, M_{\ss KK} \, \hat h \quad \, \quad  
k= M_{\ss KK} \hat k \quad , \quad 
\tilde \rho = \bar \lambda \, M^3_{\ss KK} \, \hat \rho \,, \label{dimless} 
\end{equation}
with $\bar\lambda = \lambda/(27\pi)$.
Then equations ~\eqref{equationf} and~\eqref{hequation} can be written as
\begin{equation}
\label{hatfequation}
K_z \partial_z \hat f = \hat \rho - \frac{\hat \alpha}{2} \hat k
\hat h^2 \, ,
\end{equation}
\begin{equation}
\label{hathequation}
K_z \partial_z \left [ K_z \partial_z \hat h \right ] + \left [ \hat \alpha \hat k \hat \rho - K_z^{2/3} \hat k^2 \right ] \hat h 
- \frac{\hat \alpha^2}{2 } \hat k^2 \hat h^3 = 0 \,.
\end{equation}

To determine the physical meaning of the integration constant $\hat \rho$, we
recall that the charge density of the dual axial current $J^A_{\mu}$ is computed
using the AdS/CFT dictionary as
\begin{equation}
\rho_{A}\equiv \langle J^A_0 \rangle_{\pm} \sim \lim_{z\rightarrow \infty} z^2 \partial_z {\cal A}_0 + \lim_{z\rightarrow - \infty} z^2 \partial_z {\cal A}_0\,.
\end{equation}
From the asymptotic behaviour $\hat h(z)\sim \alpha_0/z + \cdots$, which can be obtained by
perturbatively solving~\eqref{hathequation}, one deduces that the function $\hat h(z)$ does not contribute to the charge so 
$\hat \rho$ is indeed the charge density of the dual axial current (up to a constant). 
However, this does not mean that the charge density is the
same as in the homogeneous case. The reason is that the chemical
potential is determined via
\begin{equation} 
\mu_A = \frac{1}{2} (\lim_{z\rightarrow \infty} {\cal A}_0 -
\lim_{z\rightarrow -\infty} {\cal A}_0) 
= \frac{1}{2} \int_{-\infty}^{\infty}\!  {\rm d}z\, \partial_z {\cal A}_0(z) \, .
\end{equation}
Therefore, we see that the constant $\alpha_0$ \emph{does} contribute to
the value of the chemical potential, and hence modifies it in a
nontrivial way so as to give a new relation between the chemical
potential~$\mu_A$ and axial charge~$\rho_A$. We will see this
explicitly in the next section (see figure~\ref{chemicalpot}).

\subsection{Analysis of the ground state}
  
In order to find the spatially modulated phase~\eqref{fandAsoln} we
need to solve equation~\eqref{hequation} for $h(z)$ and then use this
solution when determining $f(z)$ using equation~\eqref{equationf}.
Equation~\eqref{hequation} can be solved numerically using a shooting
method.
\begin{figure}[t]
\begin{center}
\includegraphics[width=.7\textwidth]{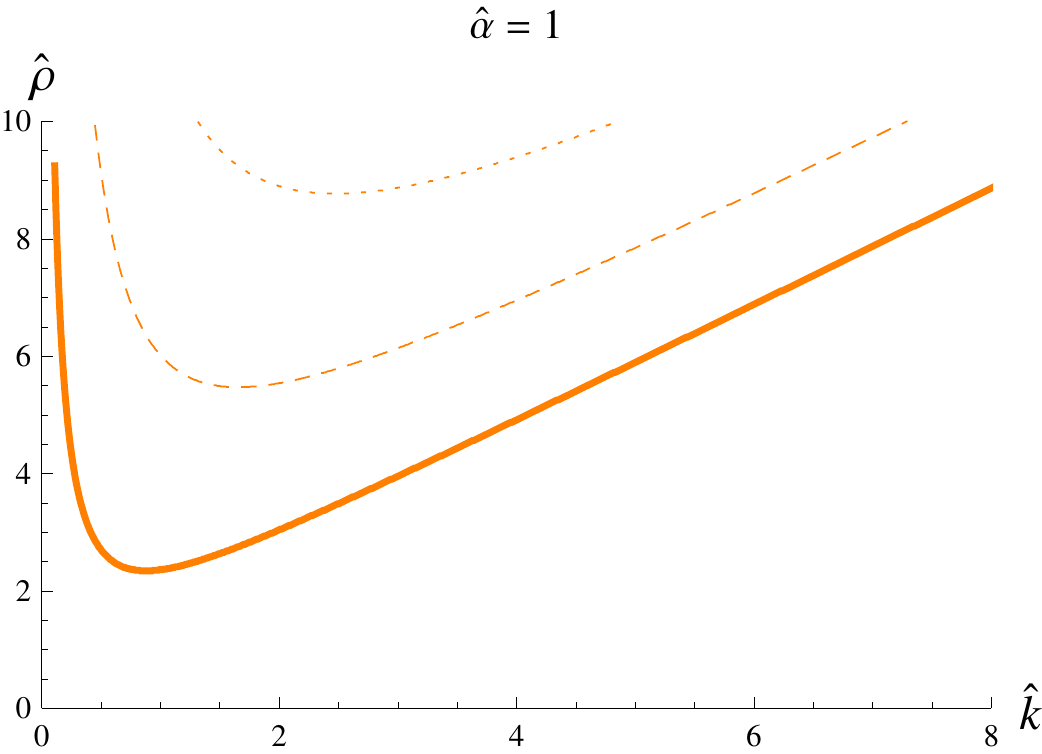}
\end{center}
\caption{Result of the stability analysis in the Maxwell-Chern-Simons
  system. The curve shows the values of $\hat\rho$ and $\hat{k}$ for
  which an instability sets in. From this analysis one finds
  $\rho_{\text{crit}} \approx 2.35$, and the corresponding mode is
  vectorial. The dashed and dotted curves correspond to
  additional instabilities of axial and vectorial nature
  respectively, which set in at higher values of the density.\label{f:RhovskMaxwell}}
\end{figure}

Before presenting our numerical solutions, let us note that
equations~\eqref{hequation} and~\eqref{equationf} have three free
parameters: the Chern-Simons coupling $\alpha$ defined
in~\eqref{e:SCS}, the wave number $k$ which sets the frequency of the
spatial modulation, and the charge density of the dual current
$\tilde{\rho}$ which appears as an integration constant in
equation~\eqref{equationf}. The wave number $k$ will be kept free,
i.e.~we solve the equation for various values of this parameter. In
string theory the constant $\alpha$ has a fixed value, but one might
consider a more phenomenological attitude in which $\alpha$ is allowed
to take on any value. In this section we will set $\alpha$ to its
string theory value, so that $\hat{\alpha} = 1$. The reason is that in
Maxwell-Chern-Simons theory, which we discuss in this section, the
actual value of coupling is not really relevant: for any $\alpha$, a
non-homogeneous state can be made to appear since one can always make
the chemical potential large enough. The situation is however different
when dealing with the DBI action, which we discuss in \S\ref{s:DBI}.

Because we do not want to introduce any external sources in the
boundary theory, we are looking for \emph{normalisable} solutions of
the equation (\ref{hequation}).  This means that, even though
(\ref{hequation}) is a second order equation, there is only one
undetermined parameter, $\alpha_0$, which governs the behaviour of
the function $h(z)$ near infinity. By expanding the equation
(\ref{hequation}) around $\pm \infty$ it is easy to see that $h$
behaves as \mbox{$h(z)\sim \alpha_0/z + \cdots$}. Using the AdS/CFT dictionary we
see that the constant $\alpha_0$ is related to the expectation value of the
dual current in the directions $\vec{\eta}_1$ and $\vec{\eta}_2$ on
the boundary,
\begin{equation}
\label{hoexpression}
\begin{aligned}
\langle J_{1}  \rangle &=   \lim_{z\rightarrow  \infty} z^2 {\cal A}_{\vec{\eta}_1} = \alpha_0 \cos(\vec{k} \cdot \vec{x}) \,,\\[1ex]
\langle J_{2} \rangle &= \lim_{z\rightarrow  \infty} z^2
{\cal A}_{\vec{\eta}_2} =  \alpha_0 \sin(\vec{k} \cdot \vec{x})\, .
\end{aligned}
\end{equation}
If the current is vectorial, then the same values are obtained by taking
the limit $z\rightarrow \pm \infty$, otherwise it is axial.

\begin{figure}
\begin{center}
\includegraphics[width=.49\textwidth]{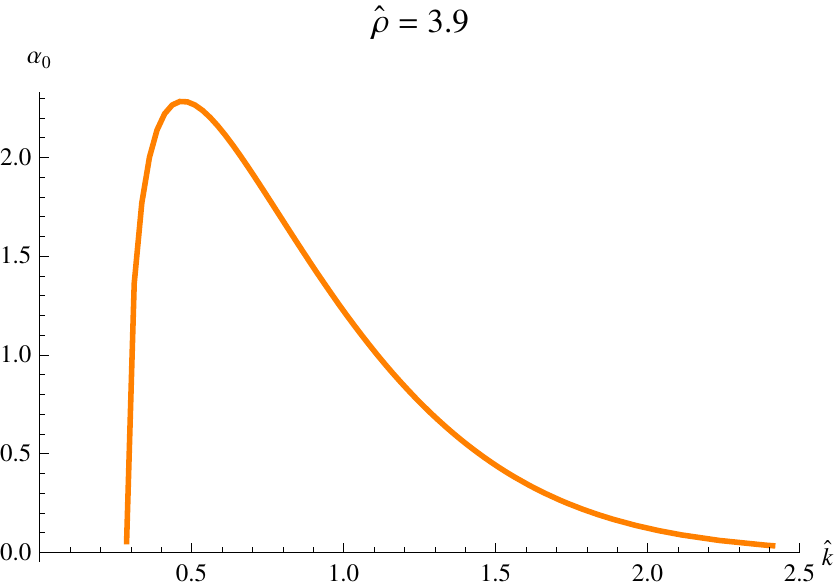}
\raisebox{2ex}{\includegraphics[width=.49\textwidth]{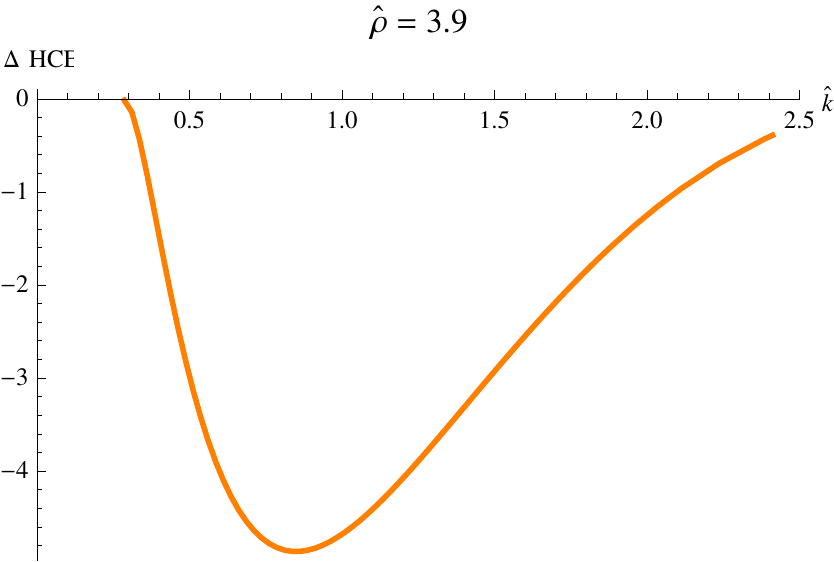}}
\end{center}
\caption{The charge density of the non-homogeneous solution, as well as
  the energy difference with respect to the homogeneous vacuum, as a
  function of the momentum. These plots show the range of momentum
  values ($k_{\text{min}},k_{\text{max}}$) for which non-homogeneous solutions
  exist. Both plots are made for a charge density $\hat{\rho}=3.9$.\label{kminkmaxs}}.
\end{figure}

Next, we use a shooting method to numerically solve (\ref{hequation})
for various values of charge densities $\rho$ and momenta $k$. The
first observation is that there is a critical value of the charge
density (and corresponding potential) so that if the density is larger
than $\rho_{\text{crit}}$, non-homogeneous solutions will exist. This
critical value is best obtained using a linear analysis, looking for
unstable modes in the homogeneous vacuum. A fluctuation with frequency
$\omega$ satisfies the linear truncation of~\eqref{hequation} with an
additional term proportional to $\omega$ added. A marginally unstable
mode is then obtained by solving for modes which have $\omega=0$,
i.e.~modes satisfying the linear truncation of~\eqref{hequation}. Such
modes occur for one-dimensional subspaces of the $\rho-k$
plane. Figure~\ref{f:RhovskMaxwell} depicts the result of this
analysis, which shows that there are various branches of unstable
modes. More specifically, we have found that for any $\hat{\rho} >
\hat{\rho}_{\text{crit}} \approx 2.35$, instabilities exist. 

Corresponding to these instabilities are new ground states. For any
momentum in the range
$[\hat{k}_{\text{min}}(\hat\rho),\hat{k}_{\text{max}}(\hat\rho)]$,
where $\hat{k}_{\text{min}}$ and $\hat{k}_{\text{max}}$ are the two
values of $\hat{k}$ on a line of instability at fixed $\hat\rho$, we
find non-homogeneous solutions of the non-linear equations. See figure
(\ref{kminkmaxs}) for an example at $\hat\rho=3.9$. However, among all
solutions for a given $\hat\rho$, there is only one momentum
$\hat{k}=\hat{k}_{\text{gs}}(\hat\rho)$ for which the solution has
minimal energy. Hence this value specifies the unique ground state for
a given charge density or chemical potential.  We should emphasise
that when we determine the ground state, we have compared the
homogeneous and non-homogeneous solutions at \emph{fixed charge
  density} $\hat{\rho}$. In other words we are working in the
\emph{canonical ensemble}. One could also work in the grand canonical
ensemble, comparing solutions at the same value of the chemical
potential~$\hat{\mu}_A$. The results which are obtained in that case
are qualitatively the same as in the canonical ensemble, so we choose
to present only the canonical ensemble results here.

\begin{figure}
\begin{center}
\includegraphics[width=.7\textwidth]{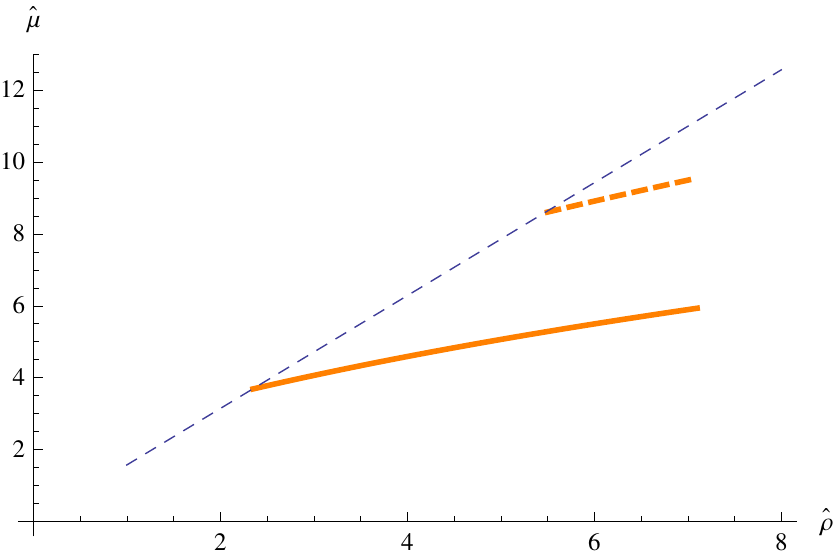}
\end{center}
\caption{The chemical potential as a function of the charge density
  for the non-homogeneous ground state. The thick dashed line
  describes the higher-energy, axial branch of the solutions, which
  are meta-stable. The thin diagonal dashed line shows the relation
  $\hat\mu_A = \pi \hat\rho_A /2$ for the homogeneous vacuum.\label{chemicalpot}}
\end{figure}

\begin{figure}
\begin{center}
\includegraphics[width=.7\textwidth]{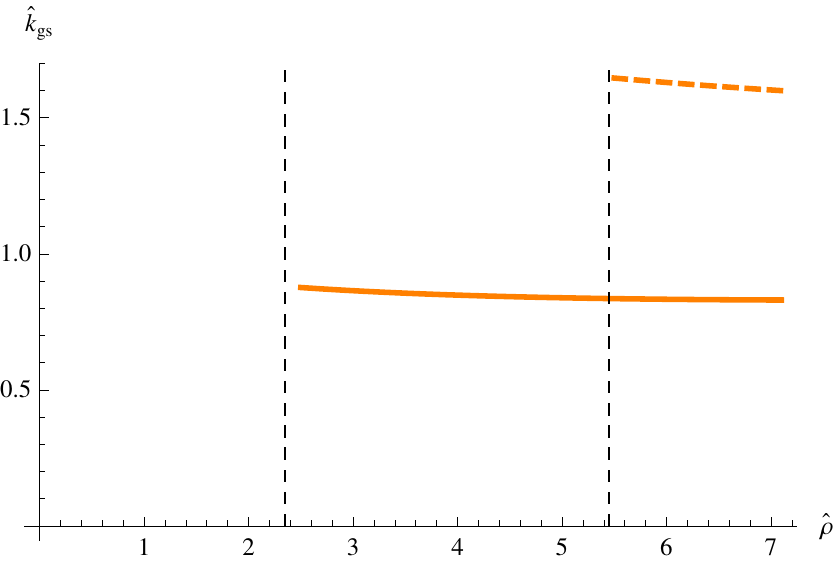}
\end{center}
\caption{The momentum of the spatial modulation as a function of the
  charge density for the non-homogeneous ground state. The dashed line
  describes the higher energy, axial branch. Observe that the momentum
  is practically independent of the density. \label{kgsofrho}}
\end{figure}

\begin{figure}
\begin{center}
\includegraphics[width=.7\textwidth]{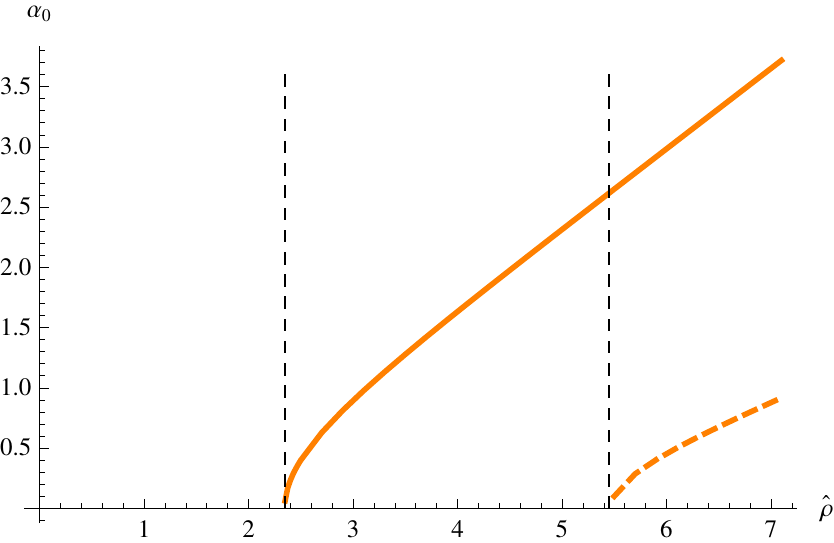}
\end{center}
\caption{The amplitude of the expectation value of the currents
  $\langle J_{\eta_1} \rangle$ or $\langle J_{\eta_2} \rangle$ as a
  function of the density. The dashed curve is again the unstable
  axial branch.\label{Jvalues}}
\end{figure}

The solutions describing ground state are uniquely specified by the value
of the charge density $\hat{\rho}$. In other words, the parameters
$(\alpha_0,\hat{k}_{\text{gs}},\hat{\mu}_A)$ specifying the ground state are all
functions of the charge density $\hat{\rho}$. The dependence of the chemical
potential on the charge density is plotted in
figure~\ref{chemicalpot}, which also shows the relation between these
two parameters in the homogeneous solution. We see that the relation
is again (almost) linear in the non-homogeneous case, but for a given
value of the charge density the chemical potential in the homogeneous case
is larger than in the non-homogeneous case.  The dependence of the
momentum $\hat{k}_{\text{gs}}$ on the density $\hat{\rho}$ is shown in
figure~\ref{kgsofrho}. Interestingly, we see that the momentum is almost
independent of the charge density. It would be interesting to
understand better the reason for this behaviour. To see if this is a
consequence of the Maxwell approximation of the DBI action, we will examine
this relation again in the nonlinear DBI case in the next section.
Finally, the dependence of the constant $\alpha_0$ on the density $\hat\rho$ is
shown in figure~\ref{Jvalues}. As is visible from
equation~\eqref{hoexpression} this constant is equal to the amplitude
of the current components $J_{\eta_1}$ and $J_{\eta_2}$.


\begin{figure}
\begin{center}
\includegraphics[width=.7\textwidth]{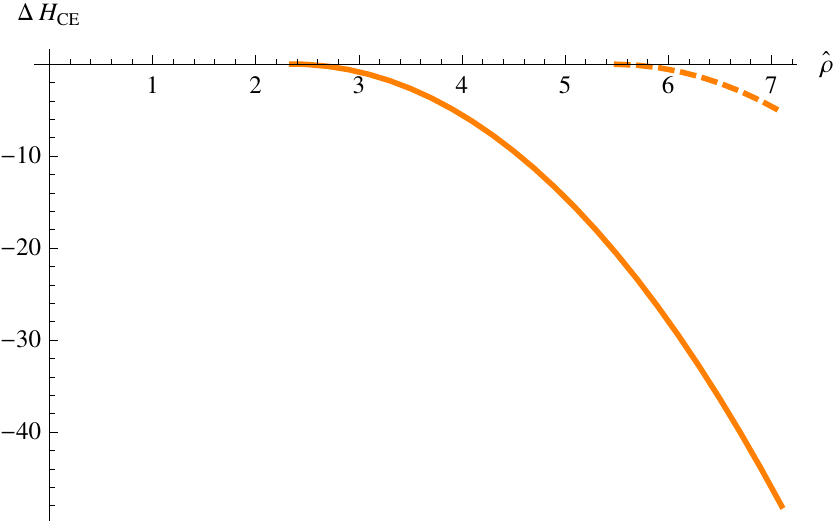}
\end{center}
\caption{The difference between the free energies of the homogeneous and
  non-homogeneous configurations as a function of the charge
  density. The dashed line depicts the axial ground state, which
  clearly has higher energy and is therefore unstable. \label{Energydiff}}
\end{figure}

The difference between the free energies of the homogeneous and
non-homo\-ge\-neous configurations (see equation~\eqref{Hbulkfull}) for
fixed value of the charge density $\rho$ is displayed in
figure~\ref{Energydiff}. As required, we see that when $\hat\rho >
\hat\rho_{\text{crit}}$, the free energy of the non-homogeneous
configuration is lower than the one of the homogeneous configuration,
indicating that the system will settle in the new ground state.
We also see that for large enough charge density, $\hat\rho > 5.4$, a new
branch of solutions appears. While the first non-homogeneous branch is
given by an even function $h(z)$, thus describing a \emph{vectorial}
current, the next branch is an odd function, describing an
\emph{axial} current. As the chemical potential is increased further,
yet more branches occur, alternating between vectorial and axial.
However, as is manifest from figure~\ref{Energydiff}, all higher
branches have higher values of the free energy, and thus correspond
to excited states, not ground states.

Finally we should note that while the ground state is characterised
by \emph{vectorial} spatial components $\eta_1$ and $\eta_2$ of the
current $J$, the charge density is always axial. This is a consequence
of the fact that the integrand in~\eqref{equationf} is an even function for
any function $h(z)$.

In conclusion, we see that the Sakai-Sugimoto model predicts that,
even at zero temperature, a non-trivial condensate of vector mesons
forms for sufficiently large value of the axial chemical potential
$\mu_A$. This condensate is non-homogeneous, and its spatial
modulation has a momentum vector which is almost constant as a
function of the chemical potential. These results complement those in
deconfined phase obtained by~\cite{Ooguri:2010xs}.

\section{Spatially modulated phase for the DBI action}
\label{s:DBI}

In the Maxwell-Chern-Simons theory, an instability towards a new
ground state of the type found in the previous section will always
occur, independent of the strength of the Chern-Simons coupling. This
is simply because the electric field (and hence the axial chemical
potential $\mu_A$) can be made arbitrarily large. However, given the
fact that the Chern-Simons term is of order $\lambda^{-1}$ with
respect to the Maxwell term, the fields in this new ground state are
necessarily of order $\lambda$. The solution found in the previous
section can therefore not be seen as an approximate solution to the
full DBI system, since the truncation that leads to the Maxwell action
assumes that $\lambda$ is small. Re-analysing the computation for the
DBI action may thus yield qualitatively new results.\footnote{Note
  that because the momentum scale $k$ of the spatial modulation is set
  by $M_{kk}$ (as follows from \eqref{dimless}), higher-derivative
  corrections can be ignored when $l_s M_{kk}$ is sufficiently small.}  

Furthermore, in the full DBI system, there is an upper bound on the
electric field, which may prevent the instability from setting
in~\cite{Ooguri:2010xs}.  A final reason to repeat the previous
analysis for the DBI-Chern-Simons action is to examine the question
whether the non-linearities of the DBI action modify the values of the
ground state momentum $k_{\text{gs}}$, perhaps making it dependent on
the charge density.

Our starting point is the DBI action for the D8-$\overline{{\rm D}8}$
system, after integration over the $S^4$ (on which none of the fields
which are turned on depend),
\begin{equation}
 S_{\text{DBI}} = - \int\!{\rm d}^4 x {\rm d}z \, \gamma
(z) \sqrt{- E} 
\end{equation} 
where 
\begin{equation}
E = \det (E_{mn}) \quad, \quad E_{mn} = g_{mn} + \beta(z) {\cal F}_{mn} \, ,
\end{equation}
with
\begin{equation}
\begin{aligned}
\gamma (z)
&= V_{\ss S^4} \, \mu_8 \, e^{-\phi } a(z)^{5/2} b(z)^2 = \frac{ \bar
  \lambda^3 N_c}{\pi^2} K_z^{-1/3} \, , \\[1ex]
\beta (z) &= \frac{2 \pi
  \alpha'}{a(z)} = \frac{1}{2 \bar \lambda} K_z^{1/6}
\,. \label{gammabeta} 
\end{aligned}
\end{equation} 
Here $g_{mn}$ is the effective 5-d metric defined in~\eqref{gmn} and
$a(z)$ and $b(z)$ are given by
\begin{equation}
a(z) = \frac{8}{27} M_{\ss KK} R^3 K_z^{-1/6}\,, 
\quad b(z) = \frac23 M_{\ss KK} R^3 K_z^{1/6}\,.
\end{equation}
This action is coupled to the
Chern-Simons term
\begin{equation} 
S_{\text{CS}} = \frac{\alpha}{4} \, \epsilon^{\ell m
  n p q} \int\! {\rm d}^4x {\rm d}z\, {\cal A}_{\ell} {\cal F}_{mn} {\cal F}_{pq} \,.
\end{equation} 
The coupling $\alpha$ is defined to be $\alpha = \hat \alpha N_c / (24
\pi^2)$ with $\hat{\alpha}$ fixed in string theory to $\hat
\alpha=1$. However, since we are interested in investigating the
nature of the instability we will relax this condition and consider an
arbitrary $\hat \alpha$.

The action variation leads to the DBI-CS equations 
\beqa
\partial_m \left ( \frac{\gamma \beta}{2} \sqrt{-E} E^{(\ell, m)} \right) 
+ \frac34 \alpha \epsilon^{\ell m n p q} {\cal F}_{mn} {\cal F}_{pq} \,=\, 0 \,, \label{DBICSeq}
\eeqa
and the boundary term 
\begin{equation}
\delta S_{\text{bdy}} = - \int\! {\rm d}^4 x {\rm d}z\, \partial_m \left [ \left ( \frac{\gamma \beta}{2} \sqrt{-E} E^{(\ell,m)} + \alpha \epsilon^{\ell m n p q} {\cal A}_n {\cal F}_{pq} \right) 
\delta {\cal A}_\ell \right ] \,.
\end{equation}
Here we have introduced the antisymmetric tensor  $E^{(\ell,m)}= E^{\ell m} - E^{m \ell}$ where $E^{\ell m}$ is defined by  $E^{\ell m } E_{m n} = \delta^\ell_n$ .

For the ansatz (\ref{fandAsoln}), the nonzero components of   $E^{(\ell,m)}$ can be written as
\beqa
E^{(z,0)} &=& - \frac {2 \beta g^{zz} g^{00} \partial_z f }{1 + g^{zz} g^{00} \beta^2 (\partial_z f)^2 + g^{zz} g^{xx} (\partial_z h)^2 } \, ,\cr 
E^{(z,i)} &=& - \frac{ 2 \beta g^{zz} g^{xx} \partial_z A_i}{1 + g^{zz} g^{00} \beta^2 (\partial_z f)^2 + g^{zz} g^{xx} (\partial_z h)^2 } \, , \cr
E^{(1,i)} &=& - \frac{2 \beta (g^{xx})^2 \partial_1 A_i }{1 + (g^{xx})^2 \beta^2 k^2 h^2 } \,.
\eeqa
Using these results the DBI-CS equations can be written in a form which puts all
modifications due to the DBI action in a single function $Q(z)$: the
equation~\eqref{equationf} becomes
\begin{equation}
\label{feqDBI}
Q(z) K_z \partial_z \hat f = \hat \rho - \frac{\hat \alpha}{2} \hat k
\hat h^2 \, ,
\end{equation}
while the analogue of equation~\eqref{hequation} for $h(z)$ now reads
\begin{equation}
\label{heqDBI}
Q(z) K_z \partial_z \left [ Q(z) K_z \partial_z \hat h \right ] + \left [ \hat \alpha \hat k \hat \rho - K_z^{2/3} \hat k^2 \right ] \hat h 
- \frac{\hat \alpha^2}{2 } \hat k^2 \hat h^3 = 0 \,.
\end{equation}
The function $Q(z)$ itself is given by
\begin{equation}
\label{QSS}
Q(z) = \frac{ \sqrt{ 1 + \frac{1}{4 K_z }  \left[ \hat k^2 \hat h^2 + K_z^{-2/3} \left (\hat \rho - \frac{\hat \alpha}{2}  \hat k \hat h^2 \right )^2 \right ] } }
{ \sqrt{ 1 + \frac14 K_z^{1/3} (\partial_z \hat h)^2 } }
\end{equation}
and we used again the dimensionless variables introduced in~\eqref{dimless}.

\subsection{Stability analysis}

\begin{figure}[t]
\includegraphics[width=.45\textwidth]{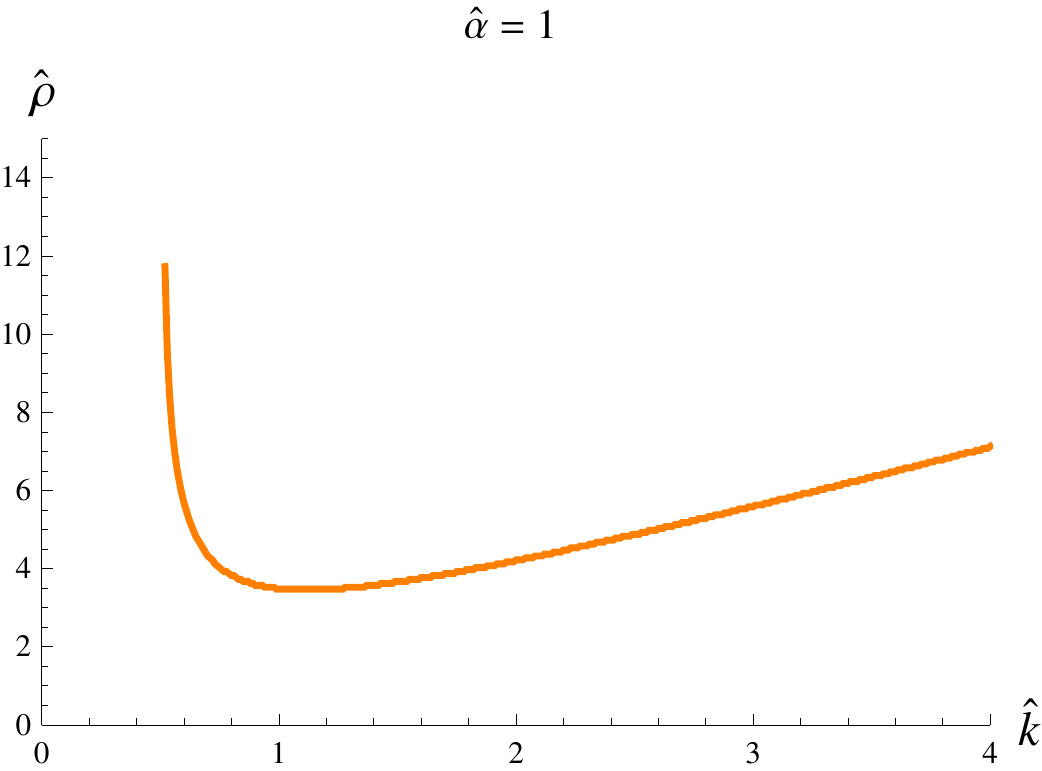}\quad
\includegraphics[width=.45\textwidth]{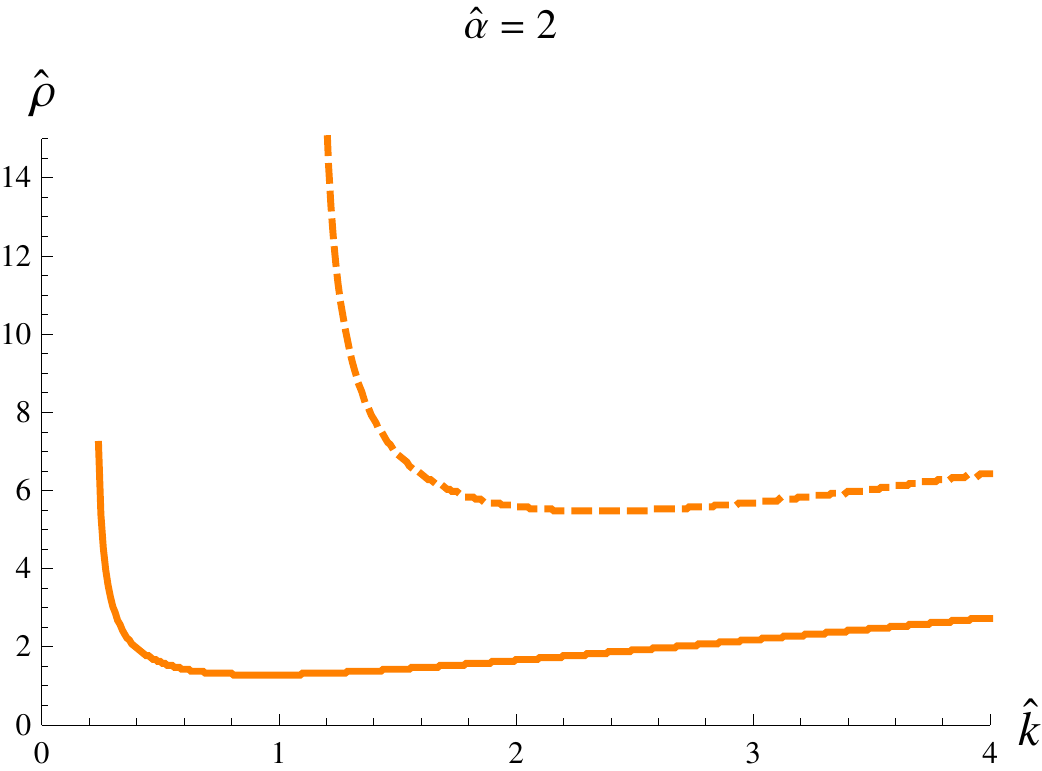}
\caption{The relation between $\rho$ and $k$ for the unstable mode of
  the DBI system, for two values of the Chern-Simons coupling:
  $\alpha=1$ for the left panel and $\alpha=2$ for the right panel. \label{f:linear_rhovsk}}
\end{figure}

Before solving the non-linear equation~\eqref{heqDBI}, it is useful to
analyse again possible instabilities in the fluctuation spectrum, in
particular in order to see how they depend on the Chern-Simons
coupling~$\alpha$. Similar to the discussion for the Maxwell case, we
can find unstable modes by solving for $h(z)$ in the linearised
version of~\eqref{heqDBI}. Results for two values of $\hat{\alpha}$
are given in figure~\ref{f:linear_rhovsk}. One observes that the
number of solutions which survive from the Maxwell truncation is
determined by the Chern-Simons coupling, and larger values typically
make the curves shift to the bottom left of the graph. For
$\hat\alpha=1$ (the value used in the Maxwell case), the critical
density is now found to be $\hat{\rho}_{\text{crit}}\approx 3.46$,
i.e.~larger than in the Maxwell case.

\begin{figure}[t]
\begin{center}
\includegraphics[width=.7\textwidth]{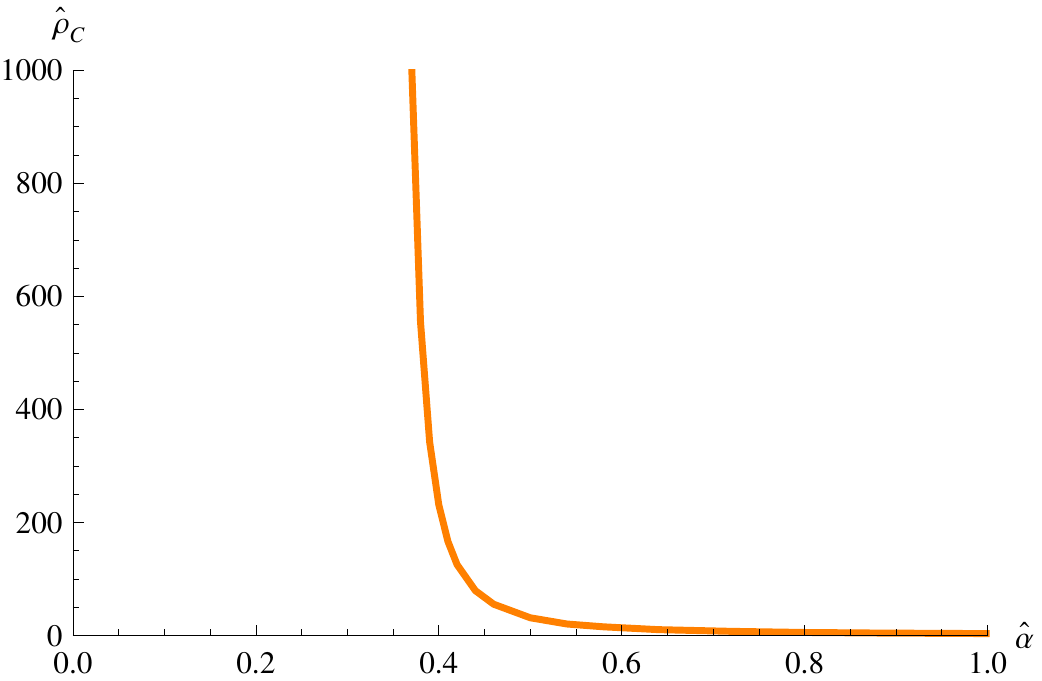}
\end{center}
\caption{The relation between the critical value of the density and
  the Chern-Simons parameter $\hat\alpha$. The critical density
  diverges as $\hat\alpha\rightarrow 1/3$.\label{f:RhocritvsalphaDBI}}
\end{figure}

From this instability analysis one can also find the minimum value of
$\hat\alpha$ for which an instability is possible at
all. Figure~\ref{f:RhocritvsalphaDBI} shows the critical density
$\hat{\rho}_{\text{crit}}$ as a function of $\hat{\alpha}$. This plot
agrees with a divergence as $\hat\alpha\rightarrow 1/3$, just as
in~\cite{Ooguri:2010xs}, indicating that there is no instability in
the system when $\hat\alpha$ takes on smaller values. For such values,
the non-linearities of the DBI action eliminate the instabilities
visible in the Maxwell truncation.

\subsection{Ground state: Maxwell versus DBI}

\begin{figure}[t]
\begin{center}
\includegraphics[width=.45\textwidth]{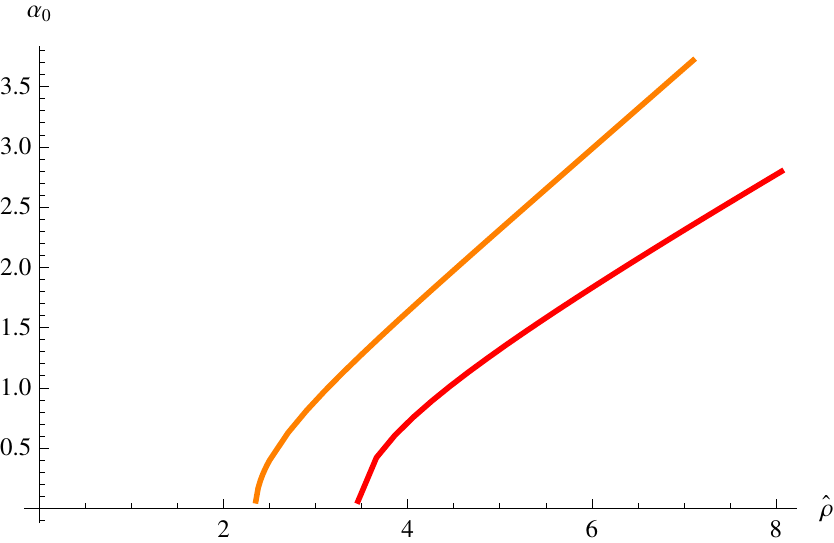}\quad
\includegraphics[width=.45\textwidth]{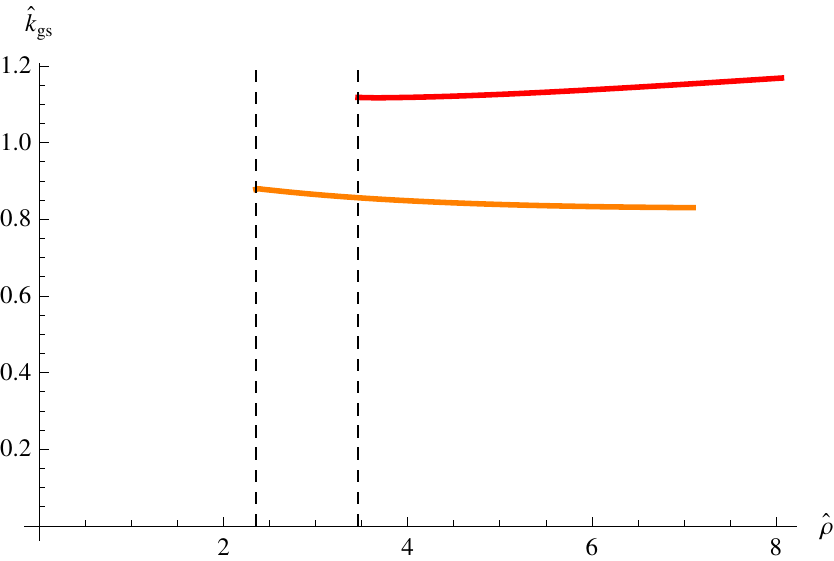}
\end{center}
\caption{Summary of the DBI analysis, with red (dark) curves depicting
  the DBI results and the Maxwell result is given in orange (light)
  for comparison. The left panel shows the expectation value of the
  current $\langle J\rangle$ as a function of the density. The
  critical density has now shifted to $\hat{\rho}_{\text{crit}}\approx
  3.46$. The right panel shows that the momentum vector is still
  practically independent of $\hat\rho$, but has also changed value.\label{f:DBInonlinear}}
\end{figure}

Finally, we present the analysis of the non-homogeneous ground state
in the DBI case. The logic of obtaining it is again the same as in the
Maxwell case, so we will not dwell on it. We have already commented on
the fact that the critical density is now substantially larger. This
is again seen explicitly in the plot of the current expectation value
$\langle J_1\rangle$ (or $\langle J_2\rangle$), given in the left
panel of figure~\ref{f:DBInonlinear}.

As far as the spatial modulation itself is concerned, we find from
this analysis that the momentum vector $k_{\text{gs}}$ for which the
non-homogeneous ground state has minimal energy is now somewhat higher
than in the Maxwell case. This is depicted in the right panel of
figure~\ref{f:DBInonlinear}. Nevertheless, it is still practically
independent of the density. 

Numerically, the instability at $\hat\rho_{\text{crit}}\approx 3.46$
corresponds, using $M_{\ss KK} = 949\,{\rm MeV}$ and $\lambda =
16.6$~\cite{Sakai:2004cn}, to a physical density of
$\rho_{\text{crit}} = 4\kappa\tilde\rho \approx 0.79\,
N_c\,\,\text{fm}^{-3}$ when $N_f=2$ (this is close to the numerical
result of~\cite{Ooguri:2010xs} up to somewhat puzzling factors of
two).

\section{Discussion and open questions}

We have analysed the effect of an axial chemical potential in the
Sakai-Sugimoto model in the low-temperature phase where chiral
symmetry is broken. We have found that for sufficiently large chemical
potential, a vectorial condensate forms, which is spatially modulated
with a momentum vector which is practically independent of the
potential or charge density. 

The results persist beyond the Maxwell truncation for the DBI action
as well. The main difference between the two is that critical density
and spatial modulation momentum take on larger values in the DBI case.
The non-linearities thus in a sense stabilise the homogeneous phase.
It would be interesting to understand if the instabilities of the type
we find persist if further non-linearities due to gravity back-reaction
are included.

It would also be interesting to understand if the new phase which we
find occurs in other QCD-like theories, and in particular in QCD
itself. The analysis of~\cite{Chernodub:2011fr} does not find this
phase, so it is worthwhile to analyse other (holographic and
non-holographic) models and isolate more carefully which feature of
the theory is responsible for the existence of the new phase. It would
also be interesting to understand how robust is our finding that the
spatial modulation is almost independent of the value of the charge
density, especially since in the large-$N_c$ analysis for the quark
chemical potential~\cite{Deryagin:1992rw}, the momentum is
proportional to the chemical potential $k_{\text{DGR}} \sim \mu_V$.

Putting together our results with those of Park and
Ooguri~\cite{Ooguri:2010xs}, we can construct a large part of the
phase diagram of the Sakai-Sugimoto model in the presence of an axial
chemical potential. We see that both in the confining and
non-confining (and chirally symmetric) phases at large enough values
of the axial chemical potential a second order phase transition to a
new non-homogeneous phase appears. To complete the phase diagram it
would be interesting to study the deconfinement in the presence of the
non-homogeneous condensate, and see if and how it modifies the
deconfinement temperature. For related work in other models see
e.g.~\cite{Ruggieri:2011xc}.

Finally, the chemical potential analysed in this paper can be embedded
in the context discussed in~\cite{Aharony:2007uu}. The question then
arises how the homogeneous (but non-isotropic) condensate found there
competes with the non-homogeneous condensate found in the present
paper. We will return to this question in a future publication.

\section*{Acknowledgements}

We thank Ofer Aharony and Cobi Sonnenschein for correspondence. This work
was sponsored in part by STFC Rolling Grant ST/G000433/1.

\vfill\eject

\section{Appendix: The Hamiltonian}

\subsection{Maxwell truncation}

Here we present the detailed formulas for the Hamiltonian associated
with the Chern-Simons action, taking special care about the surface
terms which are present for the D8-$\overline{{\rm D}8}$ system.

Our starting point is the Lagrangian density of the
Maxwell-Chern-Simons action of~\eqref{YMCS}, where we split the
space and time indices,
\begin{equation} 
{\cal L} \,=\, - \kappa\left [ {\cal
    F}^{0a}{\cal F}_{0a} + \frac12 {\cal F}^{ab} {\cal F}_{ab} \right
] \,+\, \frac{ \alpha}{4 \sqrt{- g}} \epsilon^{0 a b c d} \left [
  {\cal A}_0 {\cal F}_{ab} {\cal F}_{cd} + 4 {\cal F}_{0a} {\cal A}_b
  {\cal F}_{cd} \right ] 
\end{equation}
The conjugate momentum associated to $\partial_0 {\cal A}_a$ is given by 
\begin{equation}
 \Pi^a \,=\, \frac{ \partial
  {\cal L}}{\partial (\partial_0 {\cal A}_a)} \,=\, - 2 \kappa {\cal
  F}^{0a} \,+\, \frac{ \alpha}{ \sqrt{- g}} \epsilon^{0 a b c d} {\cal
  A}_b {\cal F}_{cd} \, , \label{momentum} 
\end{equation}
so that the Hamiltonian takes the form 
\begin{equation}
\begin{aligned} 
H_{\text{M-CS}} &= \kappa\int {\rm d}^3 \overline{x}
{\rm d}z \Big \{ \sqrt{- g} \left [ - {\cal F}^{0a} {\cal F}_{0a} + \frac12
  {\cal F}^{ab} {\cal F}_{ab} \right ] + \partial_a \left [ \sqrt{-g} \Pi^a  {\cal A}_0 \right ] \\[1ex] 
&- 2 \left [ \partial_a ( \sqrt{- g} {\cal
    F}^{a0}) + \frac{3 \alpha}{8 \kappa} \epsilon^{0 a b c d}{\cal
    F}_{ab} {\cal F}_{cd} \right ] {\cal A}_0 \Big \} \,.  
\end{aligned}
\end{equation}
The first two terms in the above Hamiltonian belong to the bulk part $H_{\text{bulk}}$, the third term is a surface term,
i.e.~a boundary Hamiltonian $H_{\text{bdy}}$, while the terms in the last line
vanish, since it is just the zeroth component of the Maxwell-Chern-Simons equations of motion.

Using the Maxwell-CS equations of motion  and the ansatz~\eqref{fandAsoln}, the canonical
momentum~\eqref{momentum} reduces to
\begin{equation}
\label{e:onshellPi}
\Pi^i = 0\,, \quad\Pi^z = \frac{2\kappa}{\sqrt{-g}} 
\left[ -\tilde{\rho} + \frac{\alpha}{2\kappa} k h^2\right]\,,
\end{equation}
and  the Hamiltonian becomes $H_{\text{M-CS}} = H_{\text{bulk}} + H_{\text{bdy}}$ with
\begin{align}
\label{Hbulkfull}
H_{\text{bulk}} &=
{\cal H} \int\! {\rm d}z\, \left[ \frac{1}{K_z} ( \hat \rho - \frac{\hat k}{2} \hat h^2  )^2 + K_z (\partial_z \hat h)^2 + K_z^{-1/3} \hat k^2 \hat h^2
  \right ]\,, \\[1ex]
\label{Hboundary}
H_{\text{bdy}} &= 2 \kappa V_x \left[\left(-\tilde{\rho} 
+ \frac{\alpha}{2\kappa} k h^2\right) f\right]_{z\rightarrow
-\infty}^{z\rightarrow \infty} = - 4 {\cal H} \,
\hat \rho \, \hat \mu_A \, , 
\end{align}
where ${\cal H} = M_{\ss KK}^4 V_x \bar \lambda^3 N_c /(8 \pi^2) $,
and we have expressed the result in terms of the dimensionless
variables defined in~\eqref{dimless}.  We hence see that for our
configuration, when there is non-vanishing value of chemical
potential, the boundary Hamiltonian $H_{\text{bdy}}$ is nonzero.

By adding to the action an extra surface term $S_{\text{bdy}}$ which for
our configuration takes the form 
\begin{equation} 
S_{\text{bdy}} \,=\, - 2 \kappa\tilde \rho \int\! {\rm d}^4 x
{\rm d}z\, \partial_z f \,,
\end{equation} 
one can remove the surface contribution to the equation of
motion~\eqref{boundaryeom} which originates from the variation of the
action~\eqref{YMCS}. This addition also simultaneously cancels the
boundary contribution to the Hamiltonian $H_{\text{bdy}}$. When
comparing free energies of the various configurations in this paper,
we therefore just need to compute the bulk contribution
$H_{\text{bulk}}$ to the Hamiltonian.

\subsection{DBI-CS Hamiltonian}

The DBI-CS Lagrangian  can be written as
\beqa
{\cal L} \,=\, - \frac{\gamma}{\sqrt{-g}} \sqrt{-E} + \frac{\alpha}{4 \sqrt{-g}} \epsilon^{\ell m n p q} {\cal A}_\ell {\cal F}_{mn} {\cal F}_{pq} \,,
\eeqa
The conjugate momentum for the gauge field is given by 
\beqa
\Pi^a = \frac{\gamma \beta }{2 \sqrt{-g}} \sqrt{-E} E^{(0,a)} + \frac{\alpha}{ \sqrt{-g}} \epsilon^{0abcd} {\cal A}_b {\cal F}_{cd} \,,
\eeqa
so the corresponding Hamiltonian is 
\beqa
H_{\text{DBI-CS}} &\,=\,& \int\! {\rm d}^3 \vec{x} {\rm d}z\, \Big \{ \gamma \sqrt{-E} \left [ 1 + \frac{\beta}{2} E^{(0,a)} {\cal F}_{0a}  \right ] + \partial_a \left [ \sqrt{-g} \Pi^a  {\cal A}_0 \right ] \cr 
&\,-\,&  \left[ \partial_a \left(  \frac{\gamma \beta }{2} \sqrt{-E} E^{(0,a)} \right ) + \frac34 \alpha \epsilon^{0 a b c d} {\cal F}_{ab}{\cal F}_{cd}  \right ] {\cal A}_0 \Big \}\,.
\eeqa
Note that the last term vanishes because it is just the $0$th component of the DBI-CS equations. 

Using the DBI-CS equations and the ansatz (\ref{fandAsoln}), the conjugate momentum reduces to
\begin{equation}
\Pi^i = 0 \quad, \quad \Pi^z = \frac{1}{\sqrt{-g}} \left ( - \bar \rho + \alpha k h^2 \right) \, ,
\end{equation}
where $\bar \rho = 2 \kappa \tilde \rho$. This is the same as the
result~\eqref{e:onshellPi} found for the Maxwell truncation. The
boundary term in the action variation takes the form
\begin{equation}
\delta S_{\text{bdy}} = \bar \rho \int\! {\rm d}^4x{\rm d}z\, \partial_z \delta f \, ,
\end{equation}
so in order to obtain a stationary action  we need to add the boundary term 
\begin{equation}
\tilde S = - \bar \rho \int\!{\rm d}^4x {\rm d}z\, \partial_z f \,.
\end{equation}
This term cancels the boundary term in the Hamiltonian. The bulk term takes the form 
\begin{equation}
H_{\text{DBI-CS}} = 8 {\cal H} \int\! {\rm d}z\, K_z^{2/3} Q(z) \left [ 1 + \frac14 K_z^{1/3} (\partial_z \hat h)^2 \right ] \, ,
\end{equation}
with $Q(z)$ given by (\ref{QSS}).

\vfill\eject


\begin{thebibliography}{14}
\expandafter\ifx\csname natexlab\endcsname\relax\def\natexlab#1{#1}\fi

\bibitem[Deryagin et~al.(1992)Deryagin, Grigoriev, and
  Rubakov]{Deryagin:1992rw}
D.~V. Deryagin, D.~Y. Grigoriev, and V.~A. Rubakov, ``Standing wave ground
  state in high density, zero temperature {QCD} at {large-$N_c$}'', {\em Int.\
  J.\ Mod.\ Phys.} {\bfseries A7} (1992)
659--681.

\bibitem[Nakamura et~al.(2010)Nakamura, Ooguri, and Park]{Nakamura:2009tf}
S.~Nakamura, H.~Ooguri, and C.-S. Park, ``Gravity dual of spatially modulated
  phase'', {\em Phys.\ Rev.} {\bfseries D81} (2010) 044018,
 \href{http://arxiv.org/abs/0911.0679}{{\ttfamily arXiv:0911.0679}}.

\bibitem[Domokos and Harvey(2007)]{Domokos:2007kt}
S.~K. Domokos and J.~A. Harvey, ``{Baryon number-induced Chern-Simons couplings
  of vector and axial-vector mesons in holographic QCD}'', {\em Phys.\ Rev.\
  Lett.} {\bfseries 99} (2007) 141602,
 \href{http://arxiv.org/abs/0704.1604}{{\ttfamily arXiv:0704.1604}}.

\bibitem[Ooguri and Park(2010)]{Ooguri:2010kt}
H.~Ooguri and C.-S. Park, ``Holographic end-point of spatially modulated phase
  transition'', {\em Phys.\ Rev.} {\bfseries D82} (2010) 126001,
 \href{http://arxiv.org/abs/1007.3737}{{\ttfamily arXiv:1007.3737}}.

\bibitem[Ooguri and Park(2011)]{Ooguri:2010xs}
H.~Ooguri and C.-S. Park, ``Spatially modulated phase in holographic
  quark-gluon plasma'', {\em Phys.\ Rev.\ Lett.} {\bfseries 106} (2011) 061601,
 \href{http://arxiv.org/abs/1011.4144}{{\ttfamily arXiv:1011.4144}}.

\bibitem[Sakai and Sugimoto(2005)]{Sakai:2004cn}
T.~Sakai and S.~Sugimoto, ``Low energy hadron physics in holographic {QCD}'',
  {\em Prog.\ Theor.\ Phys.} {\bfseries 113} (2005) 843--882,
 \href{http://arxiv.org/abs/hep-th/0412141}{{\ttfamily hep-th/0412141}}.

\bibitem[Sakai and Sugimoto(2006)]{Sakai:2005yt}
T.~Sakai and S.~Sugimoto, ``More on a holographic dual of {QCD}'', {\em Prog.\
  Theor.\ Phys.} {\bfseries 114} (2006) 1083--1118,
 \href{http://arxiv.org/abs/hep-th/0507073}{{\ttfamily hep-th/0507073}}.

\bibitem[Aharony et~al.(2007)Aharony, Peeters, Sonnenschein, and
  Zamaklar]{Aharony:2007uu}
O.~Aharony, K.~Peeters, J.~Sonnenschein, and M.~Zamaklar, ``{Rho meson
  condensation at finite isospin chemical potential in a holographic model for
  QCD}'', {\em JHEP\,} {\bfseries 082} (2007) 1007,
 \href{http://arxiv.org/abs/0709.3948}{{\ttfamily arXiv:0709.3948}}.

\bibitem[Fukushima et~al.(2008)Fukushima, Kharzeev, and
  Warringa]{Fukushima:2008xe}
K.~Fukushima, D.~E. Kharzeev, and H.~J. Warringa, ``{The Chiral Magnetic
  Effect}'', {\em Phys.\ Rev.} {\bfseries D78} (2008) 074033,
 \href{http://arxiv.org/abs/0808.3382}{{\ttfamily arXiv:0808.3382}}.

\bibitem[Kim et~al.(2010)Kim, Sahoo, and Yee]{Kim:2010pu}
K.-Y. Kim, B.~Sahoo, and H.-U. Yee, ``Holographic chiral magnetic spiral'',
  {\em JHEP\,} {\bfseries 10} (2010) 005,
 \href{http://arxiv.org/abs/1007.1985}{{\ttfamily arXiv:1007.1985}}.

\bibitem[Chuang et~al.(2011)Chuang, Dai, Kawamoto, Lin, and Yeh]{Chuang:2010ku}
W.-y. Chuang, S.-H. Dai, S.~Kawamoto, F.-L. Lin, and C.-P. Yeh, ``Dynamical
  instability of holographic {QCD} at finite density'', {\em Phys.\ Rev.}
  {\bfseries D83} (2011) 106003,
 \href{http://arxiv.org/abs/1004.0162}{{\ttfamily arXiv:1004.0162}}.

\bibitem[Chernodub and Nedelin(2011)]{Chernodub:2011fr}
M.~N. Chernodub and A.~S. Nedelin, ``Phase diagram of chirally imbalanced {QCD}
  matter'',
 \href{http://arxiv.org/abs/1102.0188}{{\ttfamily arXiv:1102.0188}}.

\bibitem[Braden et~al.(1990)Braden, Brown, Whiting, and York]{Braden:1990hw}
H.~W. Braden, J.~D. Brown, B.~F. Whiting, and J.~W. York, Jr., ``{Charged black
  hole in a grand canonical ensemble}'', {\em Phys.\ Rev.} {\bfseries D42}
  (1990)
3376--3385.

\bibitem[Ruggieri(2011)]{Ruggieri:2011xc}
M.~Ruggieri, ``The critical end point of {Quantum Chromodynamics} detected by
  chirally imbalanced quark matter'',
 \href{http://arxiv.org/abs/1103.6186}{{\ttfamily arXiv:1103.6186}}.

\end{thebibliography}

\begingroup\raggedright\endgroup

\end{document}